\documentclass[conference]{IEEEtran}
\IEEEoverridecommandlockouts
\usepackage{cite}
\usepackage{amsmath,amssymb,amsfonts}
\usepackage{algorithmic}
\usepackage{graphicx}
\usepackage{textcomp}
\usepackage{xcolor}
\usepackage{caption}
\graphicspath{ {imgs/} } 
\usepackage{multirow}
\usepackage{chemstyle}

\def\BibTeX{{\rm B\kern-.05em{\sc i\kern-.025em b}\kern-.08em
    T\kern-.1667em\lower.7ex\hbox{E}\kern-.125emX}}
\begin{document}

\title{Audio Classification of Low Feature Spectrograms Utilizing Convolutional Neural Networks}

\author{\IEEEauthorblockN{Noel Elias}
\IEEEauthorblockA{\textit{University of Texas at Austin, USA} \\
nelias@utexas.edu}
}


\maketitle

\begin{abstract}
Modern day audio signal classification techniques lack the ability to classify low feature audio signals in the form of spectrographic temporal frequency data representations. Additionally, currently utilized techniques rely on full diverse data sets that are often not representative of real-world distributions. 
This paper derives several first-of-its-kind machine learning methodologies to analyze these low feature audio spectrograms given data distributions that may have normalized, skewed, or even limited training sets.
In particular, this paper proposes several novel customized convolutional architectures to extract identifying features using binary, one-class, and siamese approaches to identify the spectrographic signature of a given audio signal. 
Utilizing these novel convolutional architectures as well as the proposed classification methods, these experiments demonstrate state-of-the-art classification accuracy and improved efficiency than traditional audio classification methods. 
\end{abstract}


\section{Introduction}
\subsection{Background}

The ongoing development of the Internet as well as the advancement of multimedia technologies have allowed for the increase of the dissipation, utilization, documentation, and creation of digital audio signals. With many different audio signals being circulated through both public and private sectors of society, the demand for tools to analyze these signals is also on the rise. In addition, with the increased utilization of big data analytics, the need for machine learning techniques to analyze these signals for signal classification, detection, and prediction are heavily sought after.  

 \begin{equation}
A_k = \sum_{n=0}^{N-1} W_N^{kn} a_n
\label{dft.eq}
\end{equation}

\begin{figure}
    \centering
    \includegraphics[width=8.3cm]{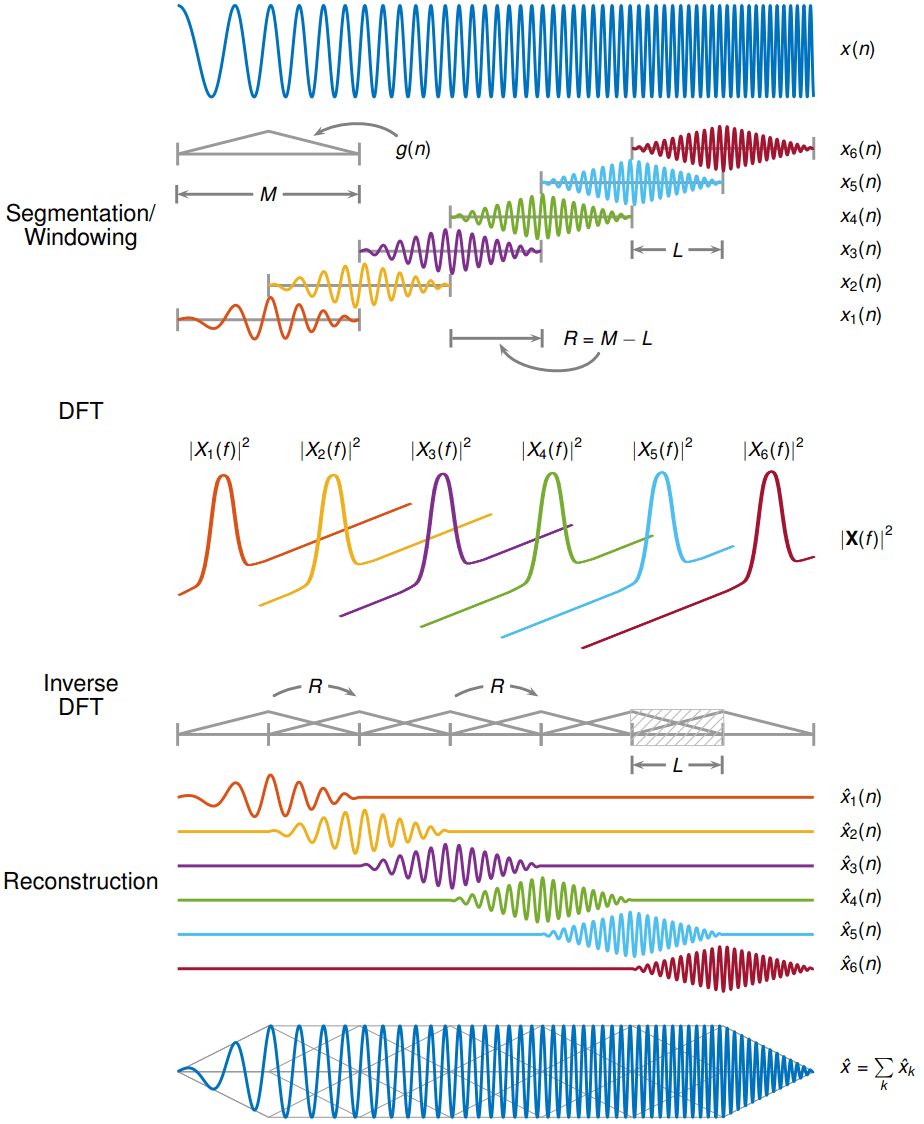}
    \caption{This illustrates algorithm to convert an audio signal to a spectrographic representation by applying the FFT on windowed portions and summing these values together. \cite{sig2spec}}
    \label{fig:fft_spectrogram}
\end{figure}

However, to arrive at a trainable representative image the digital audio signal must first be preprocessed. Most of the time, developers utilize the spectrographic form of signal which involves frequency vs time graphical representations. These spectrograms can be obtained from audio signals by simply taking their Fast Fourier Transform (FFT) of a signal(1). This is done of taking the window of the signal, summing the amplitudes of the frequencies at that duration, and then plotting/scaling these values onto a time vs frequency plot that shows the temporal relationship of different frequencies within that signal as visualized in \ref{fig:fft_spectrogram}.

Thus, we can transform audio signals into a more informative representation of spectrograms. This can then be used for a variety of analysis including detection, prediction, and classification of audio signals using machine learning.  

\subsection{Problem}
Within audio classification, most machine learning classification methods utilize spectrographic conversion alongside deep learning neural networks  to classify well-researched and recorded signals. However, the current standard models including ResNet, Inception, AlexaNet etc are unable to fill the absence of deep architectures used to classify much more common low-feature spectrographic audio signals found in nature. In essence, there are no model architectures that can be reliably used for low-feature audio signal classification for accurate and state-of-the-art signal prediction results. 

In addition, in many audio classification problems, signal training data that does not contain enough samples, data diversity, image resolution and quality, etc are unable to be properly classified with these standard deep architecture models due to unavailability of methodologies to do so.  As a result, many real-life audio data distributions are not only unclassifiable due to their low-features but are also marginalized due to their deviating training data distribution. This poses a problem not only within the audio industry but for scientific research done to understand properties of audio signals that can be exploited for human safety, security, and advancements. 
 
So, deep models and machine learning methodologies must be in place to be able to tackle these different data distributions and still provide classification techniques to predict and identify these low-feature samples. 

In this paper, we begin by proposing a variety novel convolutional neural network architectures that serve as ‘signatures’ that are able to efficiently and accurately classify previously unexplored low-feature audio signals. Also, we develop a few first-of-its-kind novel machine learning workflows for different training data sample distributions that deliver extremely accurate and high-performance predictive classification models for spectrogram data. Lastly, we compare these methodologies, discuss the results, and offer key takeaways from our evaluations.

\section{Related Work}
One of the earliest approaches to creating an audio classification model was by Guodong Guo and Stan Li \cite{guo2003content}. In this paper, the researchers developed a Support Vector Machine (SVM) model that utilized the one-against-one strategy of classifying an audio signal into a category. This worked by simply utilizing the inverted tree structure of a binary tree where each level poses as a model trained on recognizing two classes. The class that classifies the signal with a higher probability is declared the ‘winner’ and moves on to the next level until the root class is determined. The separating hyperplane or boundary of the equation for the SVMs in each round remains unchanged. The paper boasts a 11\% error rate with the SVM trained with an audio data set where signals are stored as the distance from an SVM learned boundary. 

From there, researchers have investigated utilizing end-to-end neural networks architectures (VGG, Inception,Res-Net-50, etc.) for tackling multi-class audio classification problems. Hershey, Shawn, et al \cite{hershey2017cnn} conducted an experiment showing that using graphical representations of audio signals and training these images on developed feature extraction based CNN models achieved a receiver operating characteristic score (AUC) between the ranges of 0.9 to 0.93. The ResNet model is known for learning deep features without encountering the vanishing gradient problem by utilizing ‘skip-connections’ during its training process. Another commonplace CNN model that was utilized was the AlexNet model. The 5-layer CNN model contains a combination of 3 convolutional/max pooling layers as well as a few Fully Connected layers at the end that each use the ReLu activation function. 

In fact, these previous standard model architectures including basic CNN, ResNet, VGG16, Inception, and AlexNet CNN architectures were tested against the proposed architectures and machine learning methodologies to compare their low-feature signal classification accuracy.

\section{Approach}
\subsection{Overall Design}
The overall goal of the following experiments was to propose and test machine learning approaches on low-feature spectrographic representations and different training set distributions of audio signals. 

\begin{figure}[h]
    \centering
    \includegraphics[width=4.5cm]{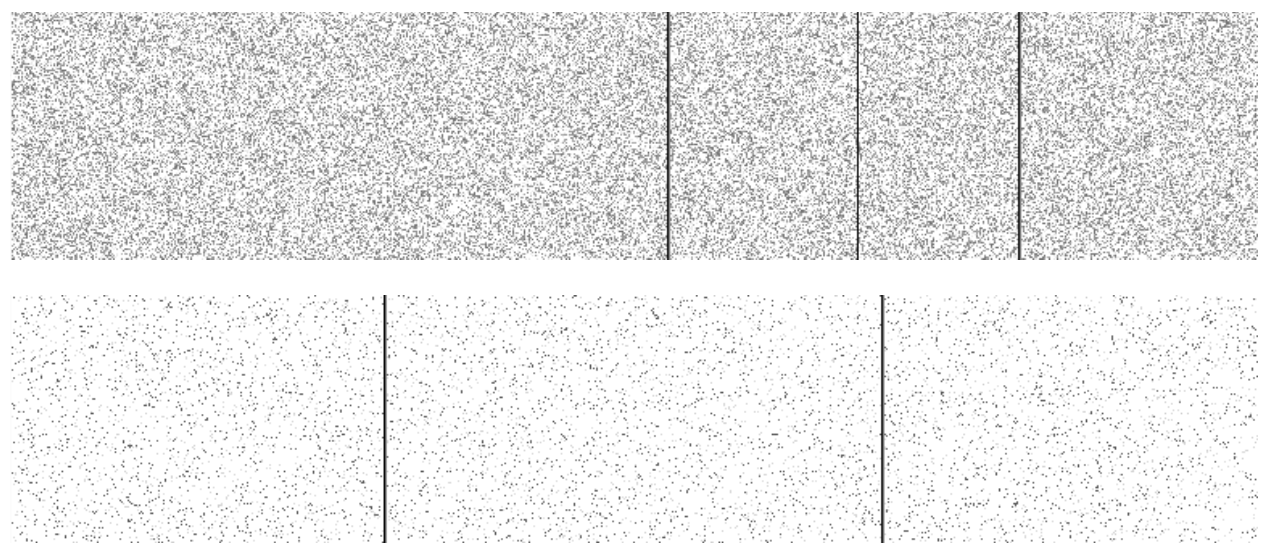}
    \caption{This illustrates the preprocessed spectrograms for sample audio signals.}
    \label{fig:spectrograms}
\end{figure}

In all the different experiments, the audio signals were generated from sine waves and contained a sample rate of 44,100 Hz. Between one to ten sine waves between 0-1000 HZ were generated to stand in for tones generated by a possible object. Alongside the tones, a random distribution of noise was normalized with the tones to resemble real-time data that was picked up by sample audio through devices like sonar, microphones, recordings etc. These audio signals were then put through the processing pipeline described in the introduction where they underwent FFT transformations through constant duration windows to result in a spectrographic representation of the temporal-frequency relationships of the signal.  In the resulting spectrograms displayed in \ref{fig:spectrograms}, the black lines represent the tonal regions surrounded by noise. Such spectrograms were crafted to closely mimic low-feature data that often also only contained few clear tones amongst noise as well. 

\subsection{Binary Classification}
Usually, audio signal training data for a certain signal contains instances of that signal as well as instances of samples with other signals or noise. For this sort of data distribution, we decided to utilize a binary approach to training classification models. 

In the binary classification approach, we utilized a one-to-all approach where the two classes we were trying to classify between were the class of the object’s signal as well as all other objects’ signals. In this approach, once we have determined the object’s signal with its unique tonal features, we generated various noise distributions to form our positive class. On the contrary, to generate all the other possible signals, we simply randomized possible tone and noise patterns. A binary classification rather than a multi-class classification set-up was used to specifically test the model's capability to classify and recognize the tested signal rather than differentiate between a cluster of signals. 

 This approach specifically deals customizing architectures with training sets of low-feature audio signal data that follow this pattern. To classify these low-feature data samples in a binary approach, the current standard CNN architectures used for audio classification were tested to see if they could classify and identify the correct signals. In addition, the LeNet model, popular for its utilization in the detention of numerical digits within the mainstream MNIST data set, was also used within this experiment. It was theorized and proposed that for a model to output distinct feature vectors that could then be learned by the different layers, we would only need a few layers of convolution. This is because spectrograms contrast for normal image data sets due to their lack of features. The only contrasts between different spectrograms are the tones between the noises which account to only a few different vertical lines. Thus, a model like the LeNet model which focuses on learning the basic vertical and horizontal features of an image by masking its noise using a few Convolutional layers followed by 3 fully connected layers was sampled in this experiment as well. 

\begin{figure*}[h]
    \centering
    \includegraphics[width=\textwidth]{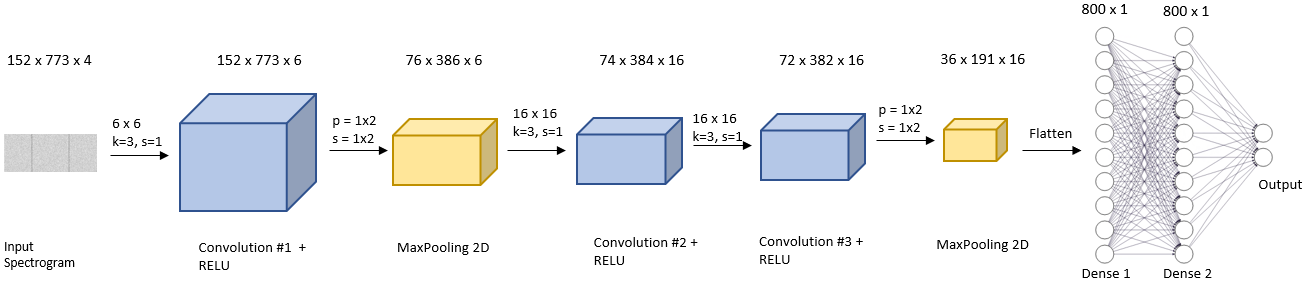}
    \caption{This illustrates the proposed Spec-CNN architecture} 
    \label{fig:specCNN}
\end{figure*}

This intuition led to the development of the novel proposed CNN architecture as shown in \ref{fig:specCNN} This architecture was called Audio Spectrogram based CNNs or Spec-CNN for short. The specific training details of the model include utilizing binary cross-entropy loss, ~20 epochs until decreasing accuracy, and a batch-size of 20 to utilize each sample within training. 

\subsection{Single Class Classification}
Similarly in many cases, only positive samples for a certain object’s audio source are available to train on with no labeled and verifiable negative samples due to computational/classification constraints. Currently, no sufficient methods exist within audio classification research that are able to correctly classify any type of audio signal let alone low feature data accurately. Often times, signals that contain this skewed data distributions are unable to be classified using any type of machine learning model. As such, the proposed methodology for this skewed training data distribution as well as the proposed CNN architecture are the first-of-its-kind within spectrographic audio classification. 

One class classification methods using machine learning are a developing field of interest within the machine learning community and are gradually being utilized in image recognition tasks.  In the single class approach, the only training data that was utilized was the spectrograms of the specified signal we wanted to be able to classify.  This paper found that the best approach for one class classification methods especially in audio classification was utilizing a deep neural network as a feature extractor and an anomaly model as a one-class classifier. As there are no one-class audio classification standards to compare this approach to, the closest comparison we could use was to test the standard deep architecture used for audio classification and see how they performed as feature extractors for one-class classification of low-feature images. The specific hyper-parameters used in this approach were to train the feature extractors utilizing categorical cross-entropy loss, ~30 epochs until decreasing accuracy, and a batch-size of 30 to utilize each sample within training.

\begin{figure*}[h]
    \centering
    \includegraphics[width=\textwidth]{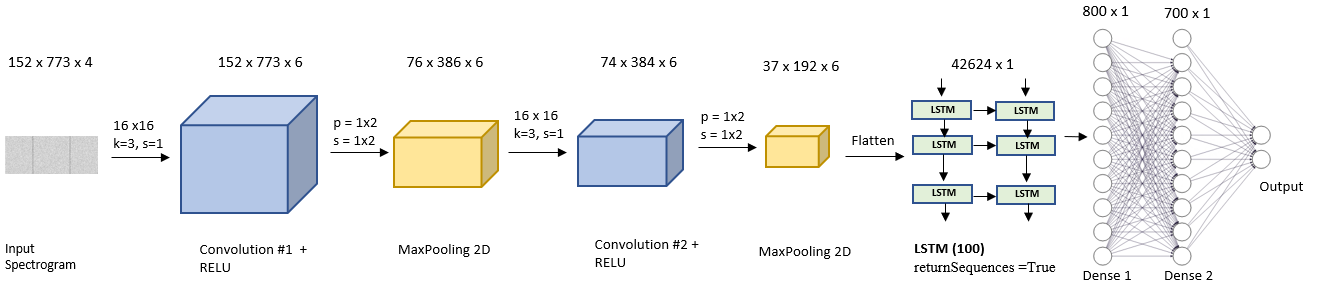}
    \caption{This illustrates the proposed OC-SpecCNN architecture} 
    \label{fig:OC-specCNN}
\end{figure*}

Within this methodology, the new CNN architecture is needed to once again be utilized for low-feature data. The proposed CNN-LSTM architecture for this feature-extraction incorporates separate Convolutional layers and LSTM layers into a sequential continuous model that utilized both these layers. This paper theorizes that due to the minimal low-features of the spectrographic data that represent the audio signals we are trying to classify, it is best to use fewer levels of convolutional  to get the features of the spectrogram. Using models that are too deep will in fact lose the low-level identifying features and result in fully connected layers that output features vectors that are too similar to differentiate between. In essence, we wanted to output a feature vector using a basic CNN model that masks the noise and outputs the locations for the tones on the spectrograms. Within the spectrogram, the tones are arranged based on appearances on a time-based scale. To contribute to this, LSTM layers were added to the results of the convolutions on the images. The memory cells of the LSTM layers allow the model to evaluate the spectrogram based on its temporal and spatial relationships to gain a better understanding of how audio signal’s different tones are related over time. This creates a much more accurate signature for any audio signal that enables more accurate classification. The methodology as well as the architecture of the CNN-LSTM or the One Class Audio Spectrogram CNN architecture (OC-SpecCNN) model is illustrated in \ref{fig:OC-specCNN}.

The proposed one-class anomaly models that were evaluated for predicting audio signals using only previous spectrograms of that class were the One-Class(OC-SVM), Gaussian Mixture Clustering, and the Random Forest Classifier Models. The SVM models that were devised and tested were One-Class SVM models that are typically used for anomaly detection (OC-SVM)\cite{yeh2009boosting}. The OC-SVM’s advantage with numerical data was novelly fitted on the spectrographic image in its vectorized form. Contrary to standard SVM models, the OC-SVM model utilizes a hypersphere boundary instead of a hyperplane boundary and trains to encompass the smallest possible hypersphere boundary containing the boundary examples. The SVM anamoly classification model utilized a 0.001 gamma, a rbf kernel, and a 0.08 nu during training.

The One Class - Random Forest Classifiers (OCRF) that were proposed for one-class classification of spectrograms of audio signals utilizes multiple decision trees to arrive at a more precise and less over-fitted prediction. Similar to a regular decision tree model, the input vector is separated to different notes based on its identifying values that eventually lead to a leaf node representing the identified class. In an OCRF, there are multiple trees that each contribute to the classification pool where the majority class is selected as the predicted class for the input. In the training, different paths are created that map to the positive class so that extraneous spectrograms map to the default “outside” class. The hyper-parametrs of this model included using 70 estimators, 0.08 contamination score, and a Isolation Forest model. 

The Gaussian Mixture model (GMM) \cite{kireeva_2021} that was also proposed and utilized in this one-class classification problem was fitted only on either the positive class features vectors or the raw spectrogram data. GMMs are basically probabilistic models for representing normally distributed clusters within large sample sets. As such, it was devised that if the GMM was fitted with positive class samples, then given a new point, we would be able to extract a log probability of whether the new point did originate from the original positive class distribution. In addition, as the GMMs outputs are log probability density function values (pdf), they were calibrated to probability scores utilizing an isotonic regression model that correlates the log probability to classifier labels.The specific training parameters for this model included a spherical covariance, 18 components, and an unbounded number of iterations.

\subsection{Siamese Network Classification}
Similarly, in some cases we may only have limited amounts of data for different audio samples due to the scarcity of the object or recording instruments themselves. In these cases, we might have previous data samples of a particular object but not enough data to train a model to identify the object’s audio signal and classify it against other samples. Currently, any audio signals include low-feature signals that have this kind of data distribution are also unclassifiable by the current machine learning standard models due to their lack of training data. To solve this pending problem, this paper proposes using another unprecedented approach of using neural network architecture known as Siamese Neural networks (SNNs) for audio signal classification \cite{koch2015siamese}. 

\begin{figure}[h]
    \centering
    \includegraphics[width=\textwidth]{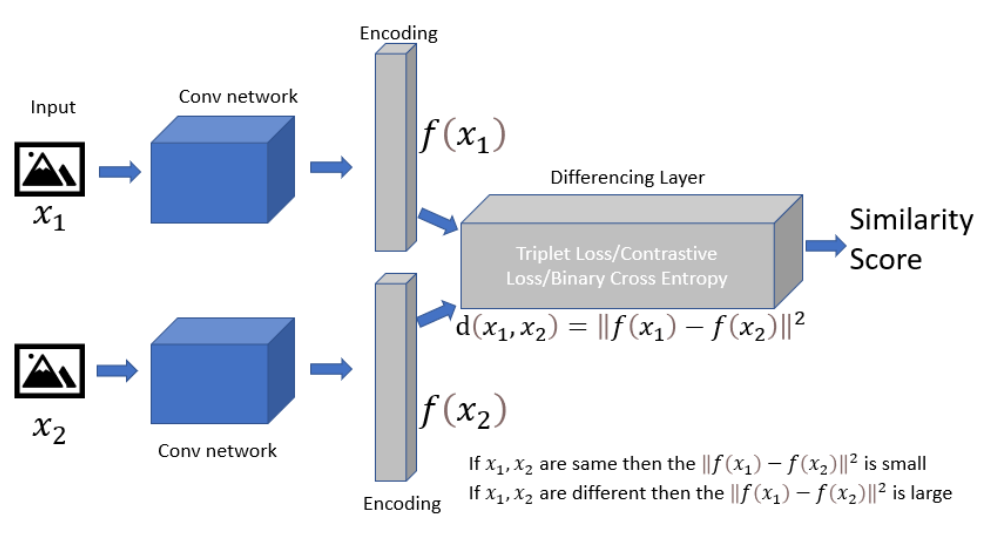 }
    \caption{Overview of Siamese network architecture. \cite{khandelwal_2021}}
    \label{fig:siamese}
\end{figure}

Siamese Neural networks \ref{fig:siamese} provide an alternative for data distributions that are constantly growing and require quick dirty classifications by providing a predictive framework for this data. In the Siamese approach we utilized a Siamese neural network architecture to input at least two different contacts we would like to compare into the neural network which outputs a similarity score of the inputs. This is done by comparing the Fully Connected layer output of two different images using some distance algorithms to estimate their similarity, and thus identify whether or not the two contacts were the same. As a result, we are able to now classify previously uninvestigated audio signals with little to no training data. Once again, due to the lack any type of Siamese standard within audio classification let alone low-feature spectrographic audio classification, the best comparison was to once again use the standard deep architectures as feature extractors for SNNs and compare the results. 

\begin{figure}[h]
    \centering
    \includegraphics[width=\textwidth]{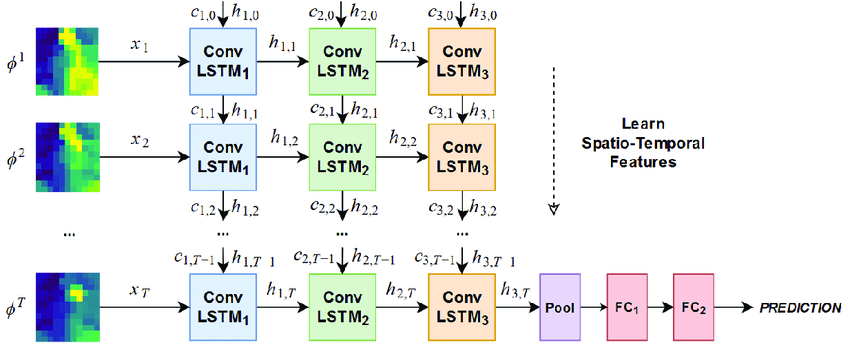}
    \caption{This represents the basic architecture of the Si-SpecCNN utilizng the ConvLSTM layers for computation \cite{zapata2019learning}}
    \label{fig:convLSTM}
\end{figure}

To enhance classification of low-feature spectrograms within this Siamese approach, a new deep architecture was also developed. The proposed ConvLSTM (Siamese-Spec-CNN or Si-SpecCNN) model contains convolutional operations within the LSTM layers as shown in \ref{fig:convLSTM}. Within the LSTM layers of the ConvLSTM, the gated channels within the LSTM undergo convolutional operations and compute Hadamard products on the 3D input data as opposed to matrix multiplication. This recurrent ConvLSTM model can be utilized to identify spatial-temporal features within spectrograms and identify dependencies between sequential data. In addition, because of the convolution operations, the additive interactions between the nodes allow for less gradient vanishing problems.

The specific training parameters for the model varied in terms of loss functions but typical ran for around 30 epochs with a batch size of 10 to accommodate memory restrains. Typically, most SNN’s utilize a contrastive loss function during the training run-time where the squared ‘distance’ between the predicted samples and the actual value is compared on both positive and negative training iterations. The model weights are only updated to either increase or decrease each feature vector from its true class. In contrast, a newer more experimental loss function known as triplet loss works by training using three inputs; an anchor, a positive and negative sample. The loss function works by manipulating the model weights to predict embedding of the positive sample closer to the anchoring image while creating contrasting embeddings of the negative sample and the anchor image.

\section{Experiments \& Results}
\subsection{Binary Classification}
As shown in Table 1, the results of these experiments proved to be promising.

\begin{figure}[h]
\centering
\resizebox{.5\linewidth}{!}{%
\begin{tabular}{ |p{2cm}||p{2cm}|p{2cm}|}
 \hline
 \multicolumn{3}{|c|}{Table 1: Binary Classification} \\
 \hline
 Feature Extractor & Accuracy & AUC (Area Under Curve)\\
 \hline
    Basic CNN   &   0.78 & 0.71\\
    \hline
    ResNet &   0.6 & 0.5 \\
    \hline
    VGG-16 &   0.89  & 0.87\\
    \hline
    Inception &   0.84  & 0.77\\
    \hline
    AlexNet &   0.88  & 0.84\\
    \hline	
    LeNet &   0.92  & 0.91\\
    \hline	
    \textbf{Proposed Spec-CNN} &   0.96  & 1.0\\
 \hline
\end{tabular}
}
\end{figure}

According to Table 1, the proposed Spec-CNN model outperformed all the standard deep architectures used for audio classification with an average AUC of 1 and average accuracy of 96 \% for classifying low-feature audio data. Thus, the Spec-CNN displayed state-of-the-art results on low-feature audio data as opposed to the current Resnet, Inception, etc standards. 

\begin{figure}[h]
    \centering
    \includegraphics[width=\textwidth, height=4.5cm]{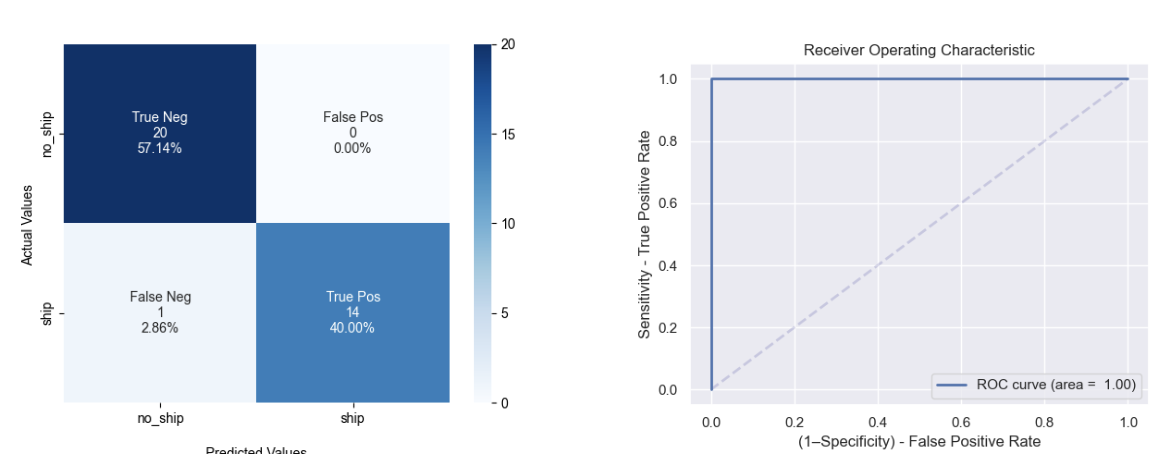}
    \caption{Shows Spec-CNN's AUC and error matrix  }
    \label{fig:binary_conf}
\end{figure}

The confusion matrix and AUC scores in \ref{fig:binary_conf}, demonstrates that using minimal epochs and a variable batch size the Spec-CNN was able to accurately differentiate and contrast the few unique tones of low-feature signals displayed on a spectrogram as opposed to all other possible signals with an astounding 96\% accuracy. 

\subsection{One-Class Classification}
The results for the one-class classification models can be illustrated as shown in Table 2.
\begin{figure}[!]
\centering
\resizebox{.5\linewidth}{!}{%
\begin{tabular}{ |p{1.5cm}|p{1.5cm}||p{1.5cm}|p{1cm}|}
 \hline
 \multicolumn{4}{|c|}{Table 2: One-Class Classification} \\
 \hline
 Feature Extractor & Classifier & Accuracy & Avg AUC\\
 \hline
 \multirow{3}{*}{Basic CNN} & SVM & 0.72 &\\ 
& RF & 0.58 & 0.48\\ 
& GMM & 0.74 & \\ 
 \hline
 \multirow{3}{*}{ResNet} & SVM & 0.33 & \\ 
& RF & 0.7 & 0.29\\ 
& GMM & 0.72 & \\ 
 \hline
  \multirow{3}{*}{VGG-16} & SVM & 0.85 & \\ 
& RF &  0.75 & 0.68\\ 
& GMM & 0.83 & \\ 
 \hline
  \multirow{3}{*}{Inception} & SVM & 0.81 & \\ 
& RF &  0.71 & 0.52\\ 
& GMM &  0.79 &\\ 
 \hline
  \multirow{3}{*}{AlexNet} & SVM & 0.82 & \\ 
& RF & 0.64 & 0.5\\ 
& GMM &  0.71 & \\ 
 \hline
  \multirow{3}{4em}{LeNet} & SVM  &0.9 & \\ 
& RF &  0.77 & 0.81\\ 
& GMM &  0.87 & \\ 
 \hline
   \multirow{3}{4em}{\textbf{Proposed OC-SpecCNN}} & SVM & 0.95 &\\ 
& RF &  0.89 & 0.93\\ 
& GMM &  0.92 &\\ 
 \hline
\end{tabular}
}
\end{figure}

Table 2 illustrates the AUC and accuracy scores of different feature vector and one-class classifier model combinations. 

Overall, the proposed CNN-LSTM combination model or the OC-SpecCNN model performed the better than the other combinations of the standard audio classification architectures used as feature extractors and anomaly classifiers scoring on average a 92\% accuracy on the training set and an avg AUC of 93\% specificity. Furthermore, the SVD classifier model with its unique bounding hyper-sphere boundary proved to be the most accurate anomaly detection model, resulting on average of 10-12\% better accuracy in the audio classification models.   

Although this approach is nuanced and there are no previous experiments or standards of one-class clasificaton of audio signals, especially low-feature signals, these experiments showed that utilizing the combining the OC-SpecCNN feature extractor as well as the SVM classifier resulted in classification accuracy that was even comparable to the binary classification standards. 

\subsection{Siamese Network Classification}
The results for the one-class classification models can be illustrated using the results of Table 3.

\begin{figure}[h]
\centering
\resizebox{.5\linewidth}{!}{%
\begin{tabular}{ |p{2cm}|p{1.5cm}||p{1.5cm}|p{1cm}|}
 \hline
 \multicolumn{4}{|c|}{Table 3: Siamese Classification} \\
 \hline
 Feature Extractor & Loss & Accuracy &  AUC\\
 \hline
 \multirow{2}{4em}{Basic CNN} 
& Contrastive & 0.72 & 0.61\\ 
& Triplet & 0.7 & 0.61\\ 
 \hline
 \multirow{2}{*}{ResNet} 
& Contrastive & 0.67 & 0.45\\ 
& Triplet & 0.56 & 0.48\\ 
 \hline
  \multirow{2}{*}{VGG-16} 
& Contrastive & 0.83 & 0.73\\ 
& Triplet & 0.76 & 0.67\\ 
 \hline
  \multirow{2}{*}{Inception} 
& Contrastive & 0.76 & 0.65\\ 
& Triplet & 0.7 & 0.6\\ 
 \hline
  \multirow{2}{*}{AlexNet} 
& Contrastive & 0.8 & 0.7\\ 
& Triplet & 0.72 & 0.59\\ 
 \hline
  \multirow{2}{*}{LeNet} 
& Contrastive & 0.89 & 0.84\\ 
& Triplet & 0.81 & 0.8\\ 
 \hline
\multirow{2}{6em}{\textbf{Proposed Si-SpecCNN}} 
& Contrastive & 0.92 & 0.91\\ 
& Triplet & 0.87 & 0.82\\ 
 \hline
\end{tabular}
}
\end{figure}
Table 3  shows evaluation of the Contrastive (C) and Triplet Loss (T) functions used for different feature extractors. 

As shown in Table 3 the most efficient siamese network model was the proposed ConvLSTM or Si-SpecCNN architecture with the best contrastive model performing with 92 \% accuracy and and average AUC score of 0.91 that outperformed all compared standards. This model utilized the contrastive loss function alongside the Euclidean distancing layer. As such, the experimented has justified the use of a Si-SpecCNN architecture for audio signal distributions especially low-feature data that may have limited samples with its almost state-of-the-art results. 

Overall, the contrastive loss function proved to be the most efficient as it was able to gain an advantage of 5-6\% over all the different model combinations. This was a result of the lack of a 'greedy' approach within triplet loss where weights remain unchanged after the disguishability between negative and positive embeddings becomes significant.

\section{Key Takeaways \& Implementation Decisions}
Here are some key decisions and experimental advances that were made:
\begin{itemize}
    \item Most feature extraction models were overfitted with many epochs in an effort to make the models memorize the images that constituted the positive class of the audio signal we were trying to identify. These helped the anomaly models create decision boundaries for the classes. 
    \item Many feature extraction models were also trained with a lower batch size in an effort to make every training example of the positive class make a greater impact on the model weights. The length of CNN stride was also prioritized over the stride height in order to maintains spectrogram formatted data. 
    \item Contrary to one would assume, the data recommended using fewer convolutional layers in order to extract the basic distinct tonal information for each spectrogram. Going much deeper, like the current deep architecture standards (Resnet, Inceptions etc), proved to over analyze the graphical data to point at which the more complex features of the spectrograms constituted of very similar embedding and drowned out the simple tonal regions of low-feature signals, resulting in many false positives. 

\end{itemize}

\section{Conclusion}

This paper's novelty lies in its proposals to solve two different important problems: the fundamental lack of accurate deep learning architectures to classify common real-world low-feature spectrographic audio signals as well as the absence of machine learning methodologies to classify everyday signals with more realistic substandard training data.  

Specifically, for data distributions with ample training data and the proposed Spec-CNN architecture performed better than all counterpart feature extractors in binary classification between the source object and various noise. In addition, in cases where only positive samples are available, the novel OC-SpecCNN architecture coupled with a OC-SVM classifier architecture performed significantly better, ~95\%, than the other standards. Lastly, for data distributions with relatively few samples the proposed Si-SpecCNN architecture with contrastive training loss proved better than the other tested methods. 

These first-of-its-kind solutions will enable audio researchers to apply these techniques to more real-life scenarios for the better understanding in fields like sonar, environmental studies, multimedia applications etc. Overall, we hope that this paper marks the start of utilizng Spec-CNNs for analyzing more real-world low-feature audio signals as well less robust but more common audio data sets for the quick, efficient, and accurate classification of audio signals in future research. 

{\small
\bibliographystyle{ieee_fullname}
\bibliography{egbib}
}

\end{document}